\documentclass[10pt,twocolumn,letterpaper]{article}

\usepackage[pagenumbers]{iccv} 
\definecolor{iccvblue}{rgb}{0.21,0.49,0.74}
\usepackage[pagebackref,breaklinks,colorlinks,allcolors=iccvblue]{hyperref}
\usepackage{indentfirst}
\def\paperID{*****} 
\def\confName{ICCV}
\def\confYear{2025}
\usepackage{fancyhdr}
\usepackage{etoolbox}

\title{Neural Object Detection for 4D-STEM: High-Throughput Sub-Pixel Electron Diffraction Pattern Recognition}

\author{
	Arda Genc$^{1}$, Ravit Silverstein$^{1,2}$\\
	$^{1}$Materials Department, University of California Santa Barbara, Santa Barbara, CA, USA\\
	$^{2}$Department of Materials Science and Engineering, University of Florida, Gainesville, FL, USA\\
	{\tt\small ardagenc@ucsb.edu, rsilverstein@ufl.edu}
}

\begin{document}
\maketitle

\begin{abstract}
	
High-throughput analysis of multidimensional transmission electron microscopy (TEM) datasets remains a significant challenge, restricting TEM's broader applicability in strategic materials research. Conventional workflows typically involve sequential, modular processing steps that necessitate extensive manual intervention and offline parameter tuning. In this work, we introduce an end-to-end post-processing framework for large-scale four-dimensional scanning transmission electron microscopy (4D-STEM) datasets, built around a highly efficient neural network-based object detection model. Central to our method is a sub-pixel accurate object center localization algorithm, which serves as the foundation for high-precision and high-throughput analysis of electron diffraction patterns. We demonstrate a strain measurement precision of 5x$10^{-4}$, quantified by the standard deviation of strain values within the strain-free Si substrate of a Si/SiGe multilayer TEM sample. Furthermore, by implementing an asynchronous, non-blocking object detection workflow, we achieve speeds exceeding 100 frames per second (fps), substantially accelerating the crystallographic phase identification and strain mapping in complex multiphase  metallic alloys. 
\end{abstract}    

\section{Introduction}
\label{sec:intro}

\begin{figure}[t]
	\centering
	\includegraphics[width=1\linewidth]{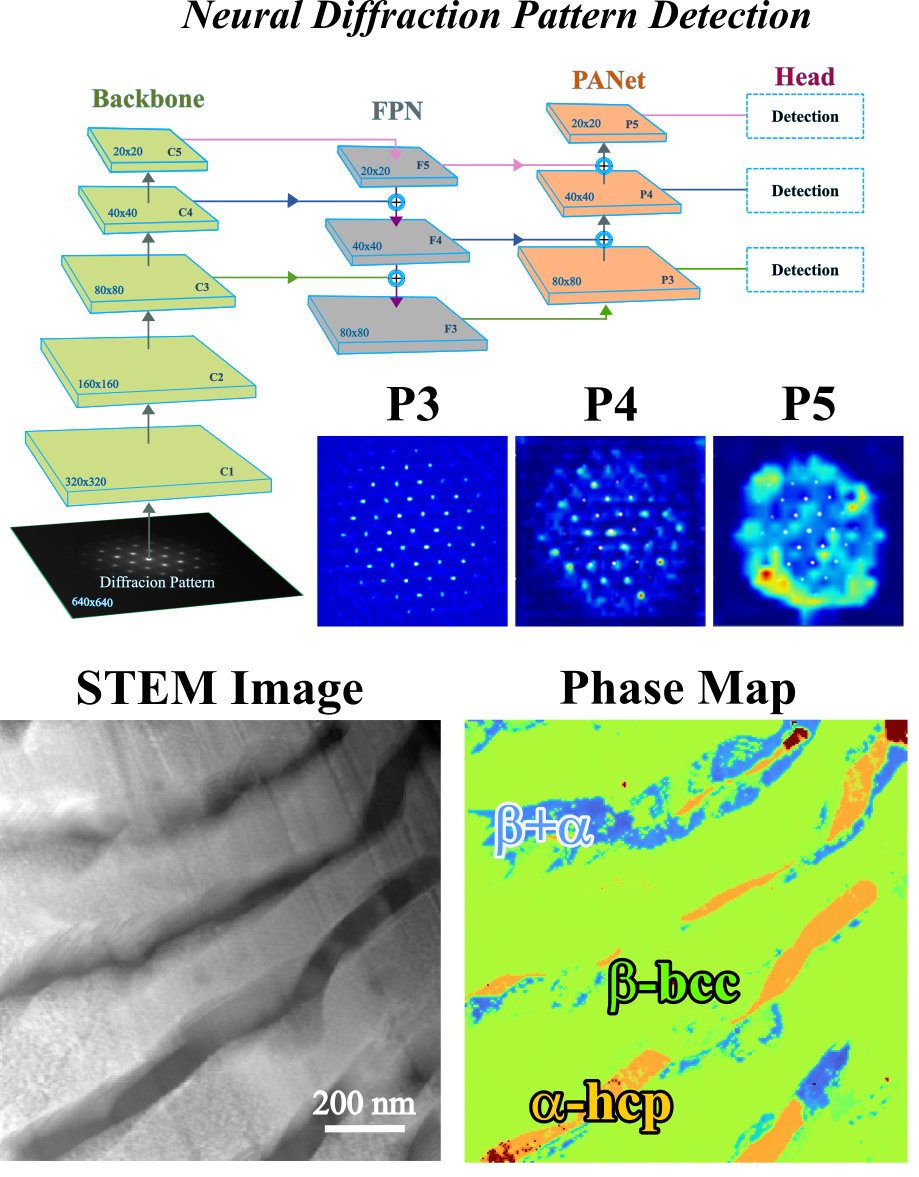}
	
	\caption{Top left to right: diagram of the neural diffraction pattern detection (NDPD) network along with exemplary activation maps from the final C2f layer, showcasing kernel responses to an electron diffraction pattern at stride levels of 8, 16, and 32. Bottom left to right: rapid phase identification of a complex phase-transformed Ti-50Nb metallic alloy by the network.}
	\label{fig:intro}
\end{figure}

Transmission electron microscopy (TEM) has emerged as an essential technique in materials science, functioning as a versatile tool for both academic and industrial applications \cite{Smith2015}. In scanning transmission electron microscopy (STEM), a focused electron probe, typically with a sub-angstrom full-width half maximum (FWHM), is scanned across a thin region of interest (ROI). The interaction of the electron beam with the material results in both elastic and inelastic scattering signals. These signals are gathered by various angular imaging detectors, offering immediate feedback on the sample’s morphology and crystallography. 

During the scanning process, imaging can be enhanced with chemical data via rapid spectrum imaging (SI) modalities. In these instances, detectors for X-ray energy-dispersive spectroscopy (XEDS) and/or electron energy-loss spectroscopy (EELS) are progressively activated to capture elemental and chemical bonding information.

In two-dimensional (2D)-STEM, a stationary diffraction pattern inherently forms during imaging; however, a diffraction pattern for every probe position is not recorded. The introduction of four-dimensional (4D)-STEM bridges this gap by enabling the acquisition and storage of each diffraction pattern using high-speed detectors, such as complementary metal-oxide-semiconductor (CMOS) or electron-pixel array detectors \cite{Tate_2016,CLOUGH20161}. 

Leveraging its broad signal collection efficiency, 4D-STEM has demonstrated record-breaking atomic resolution and facilitated numerous high-impact applications, including phase and orientation mapping, as well as the quantification of interatomic and lattice misfit displacements over a large field of view \cite{10.1017/S1431927619000497,jiang2018electron,shi2023domain,https://doi.org/10.1002/smsc.202300249,10.1017/S1431927622000101}. With the advances in high-speed cameras and automated data acquisition, it is now possible to generate hundreds of gigabytes to several terabytes of data in a single 4D-STEM session. However, adequate storage, transfer, and especially post-processing of such large datasets present a significant challenge, hindering the broader goal of achieving real-time feedback during experiments.  

Historically, electron diffraction patterns have been analyzed either through measurement from digitally recorded images or via semi-automatically using sequential pipelines involving peak-localization algorithms, such as 2D Gaussian fitting, peak pairs analysis, minimum enclosing circle, center of mass calculations, Hough transform \cite{freitag2007novel,sourty2009using,beche2013strain,yuan2019lattice,COOPER2016145,doi:10.1021/acscatal.0c00224,MAHR201538, crout2023twodimensionalstrainmappingscanning} These algorithms calculate Bragg diffraction spacings in reciprocal space or in real space using a Fourier transform. It is well-established that the precision of the position calculations by these pipelines is highly sensitive to the signal-to-noise ratio (SNR) of electron diffraction patterns, dynamical effects, variations in sample tilt and thickness \cite{li2024atomic,PEKIN2017170,park2022high,beche2013strain,COOPER2016145}.  

Here, we investigate for the first time the applicability of a fast object detection machine learning architecture in sub-pixel precision peak localization of electron diffraction patterns. We test its fidelity in detecting the position variations directly linked to the interatomic lattice distances in materials. Our objective is to develop a versatile solution to accelerate data streaming and significantly reduce the end-to-end latency of 4D-STEM data analysis, achieving improvements by several orders of magnitude.

Fig.~1 displays the main components of the neural diffraction pattern detection (NDPD) architecture \cite{Redmon2016,Jocher2023} and kernel activation responses from the network's final C2f layers. Accompanying this is a benchmark example, a phase map generated by the network from a lamellar, multiphase microstructure Ti-50Nb metallic alloy. The phase identification map reveals lamellar regions of the $\alpha$-hcp (hexagonal close-packed) phase and the $\beta$-bcc (body-centered cubic) matrix, along with overlapping zones where the lamellar phase intersects with the matrix. NDPD detects these overlapping regions by identifying double diffraction discs from two superimposed crystal structures.

The neural network's response to an example electron diffraction pattern is visualized through activation maps extracted from stride levels 8, 16, and 32, corresponding to the P3, P4, and P5 convolutional blocks within the path aggregation network (PANet). Bragg discs are identified at the finer stride level of 8 due to its high spatial resolution, while the coarser background features of the diffraction pattern are more prominent at the larger  stride levels of 16 and 32.

Our end-to-end object detection-based framework is designed to replace conventional peak-fitting and cross-correlation algorithms with a learned regression approach for direct localization of electron diffraction patterns. We explored the applicability of our object detection model for rapid and accurate analysis of large 4D-STEM datasets derived from complex material systems enhanced by an asynchronous artificial intelligence (AI) function-calling architecture.

\section{Related work}

Recently, cross-correlation methods that reference a standard diffraction disc or probe image template have been adopted for the analysis of convergent beam electron diffraction (CBED) patterns with large diffraction discs \cite{10.1017/S1431927612001274,PEKIN2017170,gammer2024strain,gammer2016local,ozdol2015strain,wang2022autodisk,10.1017/S1431927621000593,ZELTMANN2020112890}. These methods mitigate post-processing ambiguities stemming from dynamical effects in CBED patterns. Cross-correlation has proven particularly effective when using direct electron detectors with limited dynamic range, where electron dose must be regulated by distributing the CBED discs across a large number of pixels in the detector array \cite{PEKIN2017170, 10.1017/S1431927619000497}. 

In systems equipped with higher dynamic range detectors, researchers have investigated cepstral analysis of STEM nanobeam electron diffraction (NBED) patterns \cite{padgett2020exit,shao2021cepstral,https://doi.org/10.1002/smsc.202300249}. In these studies, electron diffraction patterns, often recorded at relatively shorter camera lengths and smaller electron probe  convergence angles, are transformed into the real-space cepstral domain using a logarithmic Fourier transform. Notably, these cepstrum patterns successfully disentangle interatomic distances from diffraction intensity fluctuations, which result from multiple scattering and sample mistilt \cite{COOPER2016145,padgett2020exit}. 

Machine learning (ML) has emerged as a powerful tool for the processing of 4D-STEM datasets \cite{li2019manifold,yuan2021training,shi2022uncovering,munshi2022disentangling,10.1093/micmic/ozac002,10.1017/S1431927618003422,THRONSEN2024113861}. Yoo et al \cite{yoo2024unsupervised} introduced a methodology that combines dimensionality reduction, cepstral analysis, and unsupervised ML via k-means clustering to extract interatomic distances from complex metallic microstructures. Transformation of diffraction patterns into the real-space cepstral domain effectively isolates coexisting phases and maps strain fields in a metallic shape-memory alloy. 

In the proposed pipeline taxonomy, unsupervised algorithms like k-means clustering entail nonlinear computations, such as minimizing Euclidean distance. However, they are often executed in deterministic, sequential methods \cite{https://doi.org/10.1002/smsc.202300249,yoo2024unsupervised,10.1017/S1431927621011946,SUNDE2018458,THRONSEN2024113861}, which makes them structurally similar to conventional image processing approaches, unlike the dynamic, end-to-end framework of modern deep learning models.

In a separate study, Munshi et al. \cite{munshi2022disentangling} employed a convolutional neural network (CNN)-based U-Net architecture trained on a large dataset of simulated CBED patterns to address the challenges in strain analysis arising from dynamic scattering effects, particularly in detection disc positions in thicker TEM foils \cite{munshi2022disentangling}. This approach demonstrated improved precision in strain mapping of thin films by mitigating the complexities introduced by multiple scattering and TEM sample thickness. 

However, CNN architectures, such as U-Net, often exhibit limited generalization, particularly with respect to variations in feature scale \cite{sangalli2022scale}, with model performance closely tied to the quantity and diversity of the training data \cite{Genc2022,GENC2025114116,D2NA00781A,horwathUnderstandingImportantFeatures2020,treder2023nNPipe,10.1093/micmic/ozae001,yuan2021training}. Moreover, while simulated datasets are valuable for controlled training, they are often simplistic and may fail to capture the full complexity of the experimental conditions \cite{10.1093/micmic/ozae001, treder2023nNPipe, GENC2025114116,yuan2021training}. 

Beyond deep networks like U-Net \cite{ronnebergerUNetConvolutionalNetworks2015,Genc2022}, object detection models designed with an efficient CNN architecture have transformed the pattern recognition tasks by delivering high speed and accuracy across diverse applications such as autonomous vehicles, drone surveillance, and cancer diagnosis \cite{Haifawi2023,Jia2023,Baccouche2022}. Nevertheless, their application to Bragg diffraction disc analysis in 4D-STEM datasets has yet to be explored, and the integration of these models could enhance the efficiency and accuracy of interpreting large 4D-STEM datasets. 

Rapid and accurate object localization and tracking are essential tasks in TEM analysis \cite{GENC2025114116}. Large-scale data streaming often faces bottlenecks due to I/O constraints and CPU/GPU-bound operations, resulting in high latency and reduced process efficiency. Accelerating pattern analysis with a versatile and robust framework that effectively handles electron diffraction patterns at different signal-to-noise levels, TEM camera lengths, convergence angles, and sample geometries remain a significant challenge. Developing a robust and adaptable object detection solution will enhance the application of 4D-STEM through real-time analytics.
\begin{figure*}[t]
	\centering
	\includegraphics[width=0.7\textwidth]{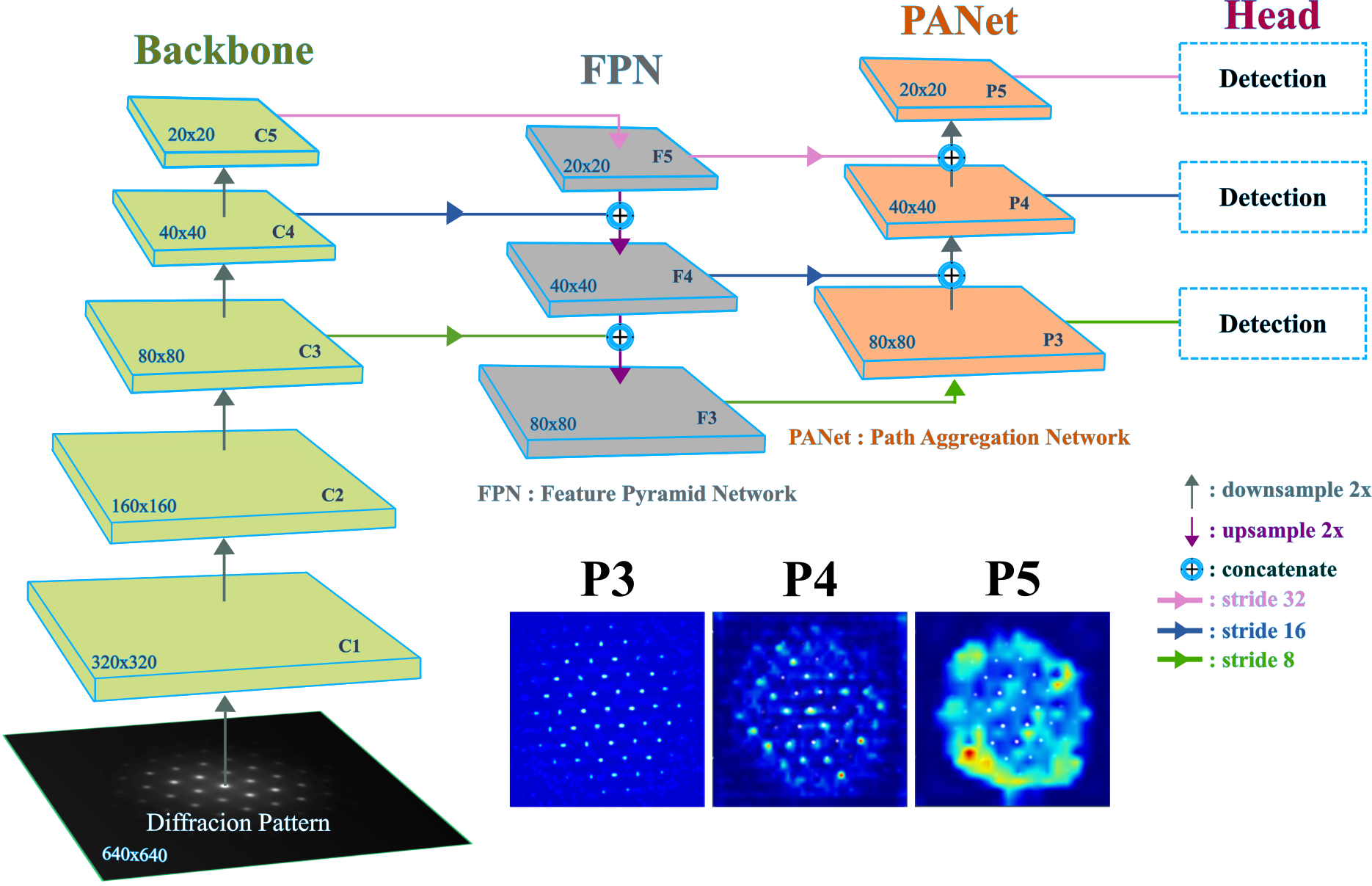}
	\caption{Illustration of the NDPD architecture, featuring the backbone (C1-C5), neck (FPN (F3-F5) and PANet (P5-P3)) components, and the detection head. }
	\label{fig:onecol_yolo}
\end{figure*} 

By adopting a non-blocking AI object detection inference framework \cite{gim2024asynchronous}, we enable real-time analysis of large 4D-STEM datasets for phase and strain mapping at speeds reaching several hundred frames per second (fps). This performance reflects a streamlined, single-step end-to-end pipeline, from inputting the 4D-STEM dataset to generating the interatomic displacement fields, all executed interactively on a standard desktop computer. The output maps are derived directly through sub-pixel position detection using NDPD. 
\section{Methodology}
\subsection{Bragg disc detection and localization}
Single-stage object detection models, such as YOLOv8, adopt a fully convolutional framework that directly predicts bounding box coordinates from feature maps \cite{Jocher2023,tian2019fcosfullyconvolutionalonestage}. This anchor box-free design enables a well-generalized \cite{GENC2025114116}, object-center-aware feature augmentation, wherein grid cells serve as reference points for predicting offsets to object centers. Such a center-based approach is particularly well-suited for the high-throughput and high-precision detection of Bragg disc centers in electron diffraction patterns \cite{Redmon2016,Jocher2023,duan2019centernetkeypointtripletsobject,tian2019fcosfullyconvolutionalonestage}. 

Fig.~2 illustrates the three main components of the YOLOv8 object detection architecture: the backbone, neck, and head. The backbone, composed of a deep CNN, is responsible for extracting hierarchical feature maps from input images. It consists of a sequence of convolutional blocks with residual connections, denoted as C1$\rightarrow$C2$\rightarrow$C3$\rightarrow$C4$\rightarrow$C5, corresponding to increasing stride levels of 2, 4, 8, 16, and 32. Lower convolutional layers preserve high-resolution spatial details like edges and textures, while higher layers capture more abstract, semantically rich features. 

The neck of the network integrates features from different stages of the backbone to generate a multi-scale representation through lateral skip-connections. This hierarchical augmentation is achieved through a combination of a feature pyramid network (FPN) \cite{Lin2017FPN} in a top-down pathway (F5$\rightarrow$F4$\rightarrow$F3) and PANet \cite{Liu2018} in a bottom-up pathway (P3$\rightarrow$P4$\rightarrow$P5). This dual-path strategy enhances the fusion of semantic and high-resolution spatial details, ensuring that each level retains strong contextual understanding and spatial precision.

The head of the network is responsible for predicting bounding boxes, objectness scores, and class probabilities at multiple spatial resolutions. By leveraging feature maps from different levels of the network, the model can detect objects across a wide range of scales.

The model predicts outputs as bounding boxes represented by the box center coordinates (x and y), along with their width and height. In YOLOv8, distribution focal loss (DFL) \cite{li2020generalized} enhances bounding box regression by modeling each side of the box (left, top, right, and bottom) as a discrete probability distribution over a set of bins. Instead of directly regressing continuous values for these parameters, DFL treats them as classification tasks over discrete bins, determined by the $reg\_max$ parameter (e.g. $reg\_max$  = 16, resulting in 17 bins). 

As shown in Fig.~3, for each parameter, the model outputs logits corresponding to each bin. These logits are mapped into a probability distribution using the softmax function. The final predicted value for each parameter is then computed as the expected value (mean) of the distribution, effectively performing a soft-argmax operation:
\begin{figure}[h]
	\centering
	\includegraphics[width=0.65\linewidth]{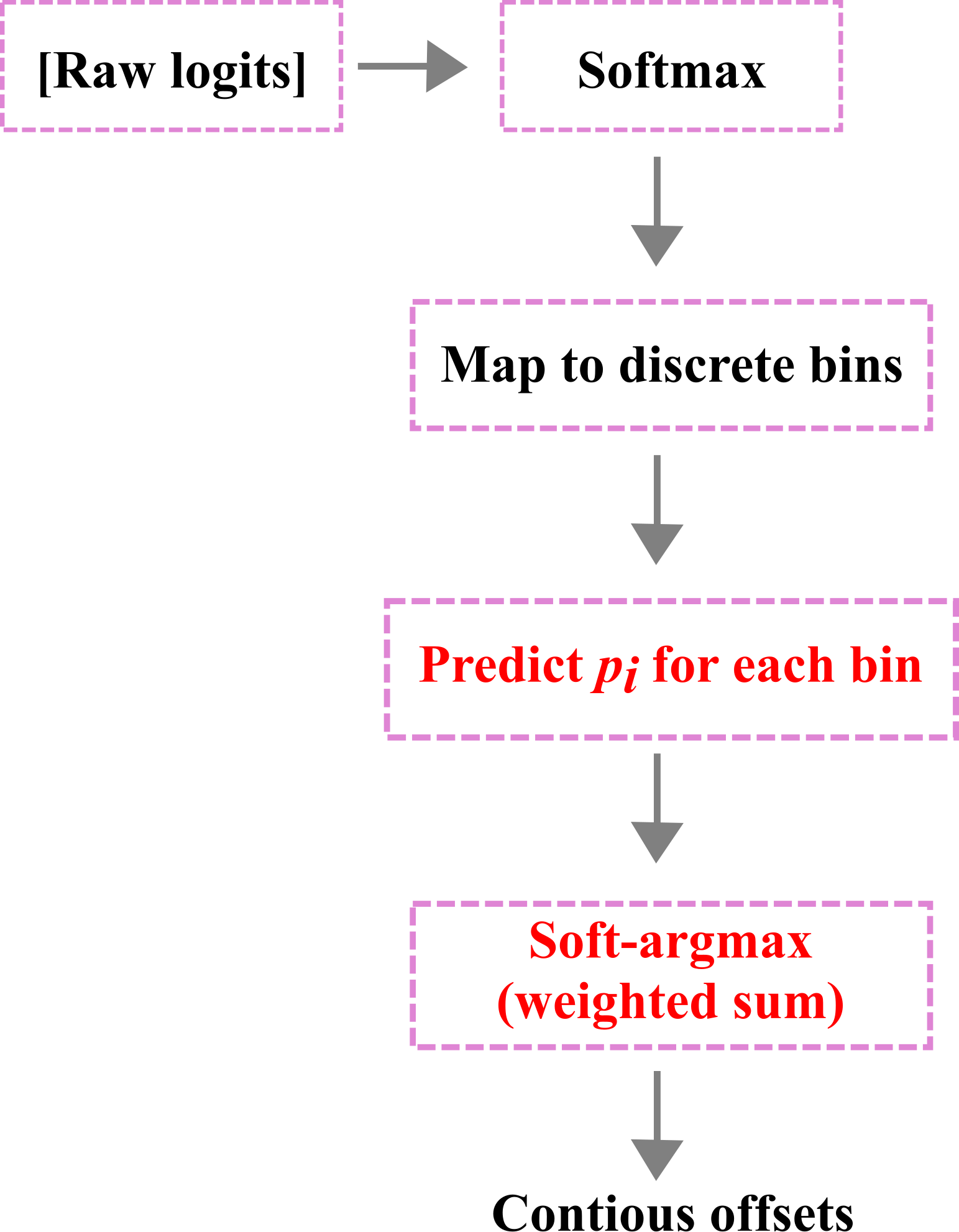}	
	\caption{The flowchart illustrates the bounding box regression process utilizing DFL.}
	\label{fig:onecol_flow}
\end{figure}

\begin{equation}
d_n = \sum_{k=0}^{\text{reg\_max}} p_k^n \cdot k
\label{eq:dn_sum}
\end{equation}
Where $p_k$ is the probability assigned to bin $k$, $d_n$ is the bounding box parameter, and $n$ is the four sides of the bounding box noted as left, top, right, and bottom.

The model predicts offsets relative to predefined anchor points located at the center of each grid cell, as illustrated in Fig.~4 for a 13x13 pixel grid overlaid on an electron diffraction disc. The final center coordinates in the input image are computed by adding the predicted offsets to the anchor point positions and are given by:
\begin{equation}
	x_{\text{center}} = \left( c_x + 0.5 + \frac{d_{\text{right}} - d_{\text{left}}}{2} \right) \cdot s
	\label{eq:x_center}
\end{equation}
\begin{equation}
	y_{\text{center}} = \left( c_y + 0.5 + \frac{d_{\text{bottom}} - d_{\text{top}}}{2} \right) \cdot s
	\label{eq:y_center}
\end{equation}
Where $c_x$ and $c_y$ are the column and row indices of the grid cell, $d_n$ is the bounding box parameter, and $s$ is the stride level.

To improve the center localization in our object detection model, we incorporated an efficient autocorrelation-based approach, leveraging the Wiener–Khinchin theorem \cite{10.1063/1.5096245,wiener1930generalized,khinchin1934korrelationstheorie}. Recently, the log-magnitude power spectrum of reciprocal-space intensities has been applied via cepstral analysis to decouple lattice periodicities in 4D-STEM diffraction patterns. Although the logarithmic operation compresses the dynamic range and suppresses dominant intensity variations, it can potentially obscure weaker signals in low SNR situations. In our implementation, spectral decomposition is performed without applying the logarithmic transform, similar to the Patterson function \cite{PhysRev.46.372}, utilizing an autocorrelation function to retain the full-intensity scaling. This strategy enhances the extraction of interatomic distances while minimizing potential ambiguities in Bragg disc intensities caused by sample tilt and thickness variations. 

The Wiener-Khinchin theorem relates the autocorrelation function of a stationary random process to its power spectral density via the inverse Fourier transform. In the quefrency domain, the autocorrelation function is given by:
\begin{equation}
	R(\Delta \vec{k}) = \mathcal{F}^{-1} \left( \left|\mathcal{F}\left[I(\vec{k}) \right] \right|^2 \right)
	\label{eq:wiener_khinchin}
\end{equation}
Here $I(\vec{k})$ is the input signal in reciprocal space, $\mathcal{F}$ denotes the Fourier transform, and $\Delta \vec{k}$ represents the displacement vector in quefrency space.

\begin{figure}[h]
	\centering
	\includegraphics[width=0.5\linewidth]{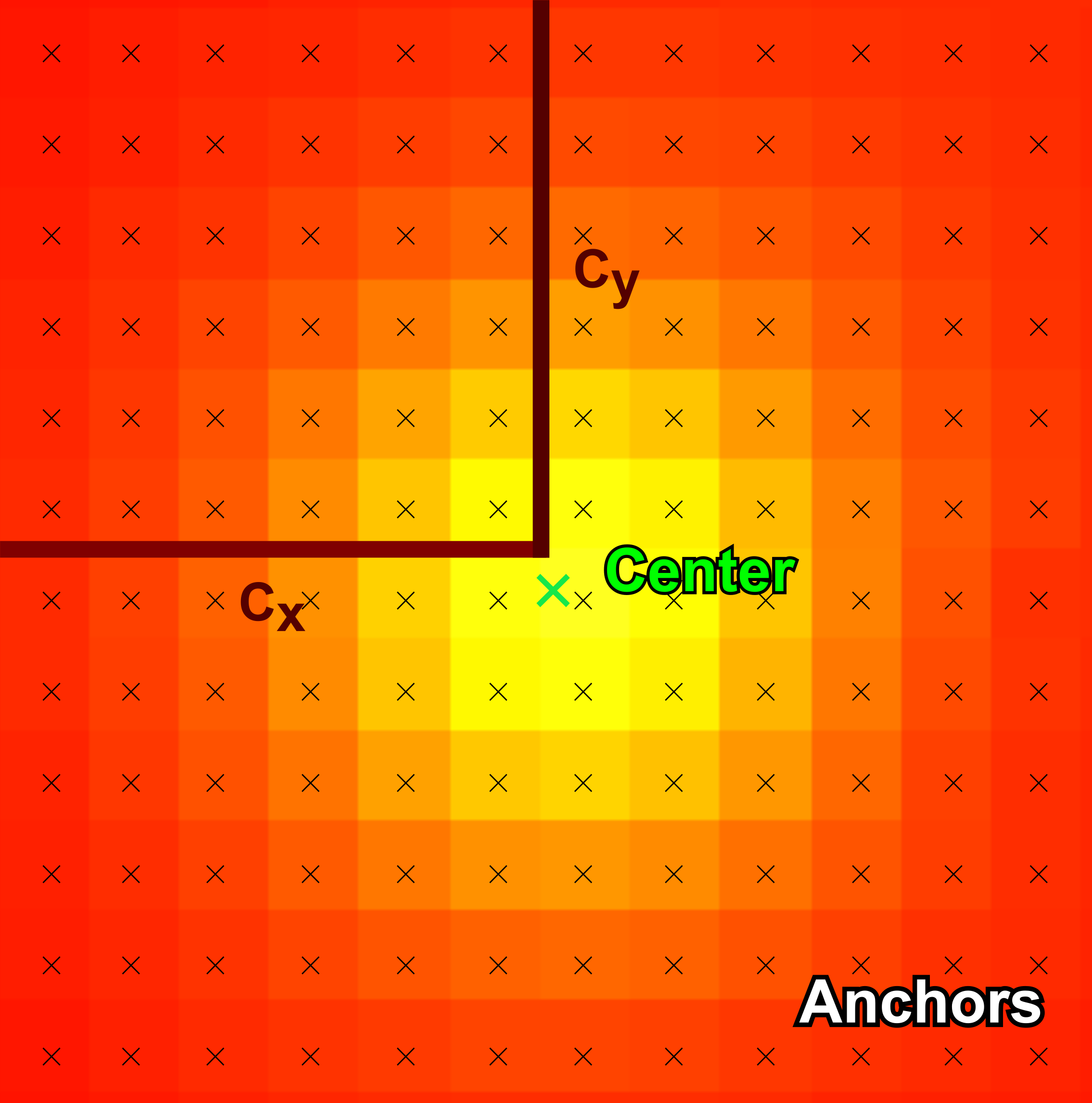}
	\caption{Illustration of anchor point-based center detection using a 13x13 pixel grid. Each anchor point corresponds to the center of a grid and is marked with a small dark cross. The predicted center of the diffraction disc is indicated by a green cross, representing the model's offset-based localization from the nearest anchor point.}
	\label{fig:onecol_anchors}
\end{figure}

An asynchronous function-calling strategy enables parallel processing of large 4D-STEM datasets \cite{gim2024asynchronous}. This design ensures that I/O operations are non-blocking during AI model inference, thereby eliminating GPU load/unload cycles and maximizing computational throughput, as shown in Fig.~5. 

To further enhance performance, we integrated automatic mixed precision (AMP) during inference \cite{micikevicius2018mixedprecisiontraining}. AMP leverages lower precision arithmetic (e.g., FP16) where appropriate, reducing memory usage and increasing computation speed without compromising model accuracy. This combination of asynchronous processing and AMP resulted in a fourfold increase in fps and a substantial reduction in end-to-end processing time.
\begin{figure}[h]
	\centering
	\includegraphics[width=0.9\linewidth]{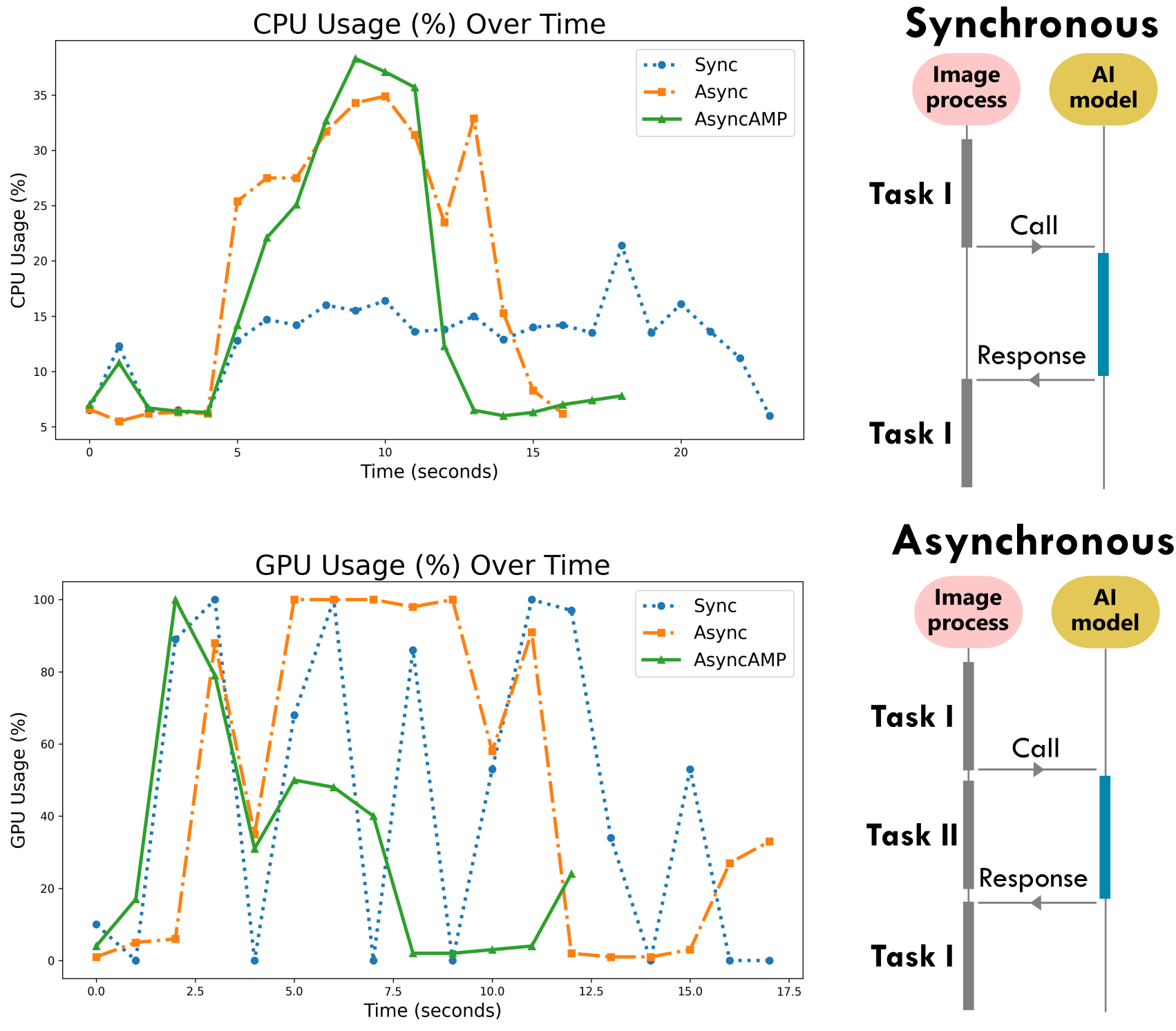}
	\caption{Comparing CPU and GPU performance in Synchronous and Asynchronous Neural Object Detection.}
	\label{fig:onecol_benchmark}
\end{figure}

\subsection{Model training and optimization}

Model training for the object detection task begins with transfer learning using the pre-trained weights of the YOLOv8 Nano (YOLOv8n) model, which serves as the foundation for the bounding box predictions. This model is then fine-tuned using manually annotated images to adapt specifically for detecting the center coordinates of the diffraction discs. Each bounding box prediction includes the center coordinates along with the box’s width and height ($b_x$, $b_y$, $b_w$, $b_h$). The YOLOv8 Nano model is designed for lightweight and efficient object detection tasks, featuring the fewest convolutional layers, channels, and parameters (3.5~million parameters). This compact design enhances the model’s ability to rapidly and accurately detect the center coordinates of diffraction discs, making it well-suited for high-throughput analysis in electron microscopy applications.

For training the object detection model, we employed mini-batch gradient descent with a batch size of 32 and a stochastic gradient descent (SGD) optimizer with weight decay regularization. The training process adhered to the default hyperparameters provided by the Ultralytics framework, which include settings for gradient averaging and the updating of first and second moments. The YOLOv8n model was initialized through transfer learning, leveraging pre-trained weights from the MS COCO (Microsoft Common Objects in Context) dataset, which comprises 330,000 annotated images across 80 object classes. For fine-tuning, we curated a dataset of 42 electron diffraction patterns obtained from a diverse range of materials, including metallic alloys and thin-film heterostructures. Each diffraction pattern was manually annotated to generate ground truth bounding boxes, and the dataset was subsequently divided into training and validation subsets to facilitate model evaluation. 

We applied rotation, vertical and horizontal flipping, brightness adjustments, and noise transformations to generate a diverse set of images capturing variations in the location and shape of the features within electron diffraction patterns. Following augmentation, our dataset consisted of 90 images with corresponding annotations for training and 12 images for validation. The YOLOv8n model was trained using an input resolution of 640x640 pixels using the computational capabilities of two NVIDIA A100 80~GB GPUs.

\section{Experiments and Results}
\subsection{Materials and datasets}
The MAG*I*CAL® sample, obtained from Ted Pella Inc, is a widely recognized and extensively studied standard in TEM for benchmarking strain measurements, primarily due to the inherent lattice mismatch between epitaxially grown silicon (Si) and silicon-germanium (SiGe) layers \cite{li2024atomic,https://doi.org/10.1002/smsc.202300249,munshi2022disentangling,beche2013strain,COOPER2016145}. This traceable standard comprises alternating layers of pure Si and SiGe alloy deposited on a single-crystal Si substrate. Specifically, it consists of five SiGe layers, each approximately 10~nm thick, alternating with pure Si layers approximately 13~nm thick, culminating in a total superlattice thickness of about 100~nm. 

We acquired XEDS-STEM profiles along the multilayer stack and measured the average composition of the SiGe layers using the Ge~K-line at 9.874~keV and Si~K-line at 1.739~keV, determining the composition to be approximately 87~at.\% Si and 13~at.\% Ge. Based on the XEDS quantification, we estimated the maximum expected out-of-plane strain value, $\varepsilon_{xx}$, to be approximately 0.85~\% using equation (5) proposed by Munshi et al. \cite{munshi2022disentangling}.
\begin{equation}
	\varepsilon_{xx} = \left( \frac{a_{\text{Ge}}}{a_{\text{Si}}} - 1 \right)(1 - x_{\text{Si}})(1 + 2\nu)
	\label{eq:strain_ge_si}
\end{equation}

Where the Ge lattice parameter is \( a_{\text{Ge}} = 0.5658~\text{nm} \), the Si lattice parameter is \( a_{\text{Si}} = 0.5431~\text{nm} \), and the Poisson ratio is \( \nu = 0.28 \).

We used a Ti-80 at.\% Nb phase-transformed metallic alloy for phase and strain field mapping. The Ti-80Nb alloy was prepared via arc melting and subsequently processed by splat quenching to produce thin foils approximately 250~µm thick. To ensure a uniform bulk composition of 80~at.\%~Nb, the foils underwent homogenization heat treatment. Oxygen enrichment was then carried out by encasing the alloy in Nb foil and sealing it in fused silica ampoules under an $\mathrm{Ar}/\mathrm{O}_{2}$ gas mixture. The sealed ampoules were heat-treated at 800°C for 2 hours, resulting in an oxygen concentration of approximately 1 at.\% within the alloy. The TEM sample was prepared by focused ion beam (FIB) milling at 30 kV using a Thermo Fisher Scientific Helios NanoLab 600. To minimize Ga$^{+}$ ion-beam induced damage, final thinning of the TEM foil was performed at 5 kV.

\subsection{Strain mapping of pseudomorphic Si/SiGe epitaxial layers}
Both cepstrum and autocorrelation-based analyses involve a trade-off between achieving high precision in peak localization and effectively averaging over multiple diffraction discs to mitigate intensity fluctuations \cite{padgett2020exit}. This balance necessitates the use of an optimum camera length during diffraction pattern acquisition, allowing the capture of a sufficient number of diffraction discs for reliable averaging while maintaining enough pixel resolution on each disc to enable accurate localization. To meet these requirements, we utilized a 512×512 pixel array and a 32-bit dynamic range GATAN Stela hybrid-pixel electron detector operated in energy-filtered mode. 

To investigate the fidelity of our NDPD workflow in strain analysis, we acquired a 4D-STEM dataset from a Si/SiGe multilayer TEM sample at an operating voltage of 200~kV. The TEM foil thickness was approximately 70~nm, as determined by EELS. The acquisition conditions were optimized so that each diffraction disc spans at least 10 pixels in diameter and extends to at least the third-order Laue zone, ensuring both spatial precision and sufficient sampling of the diffraction space.

Fig.~6 illustrates a set of diffraction patterns from the Si substrate of the Si/SiGe TEM sample. The object detection model generates bounding box predictions, and from these predictions, we extract the center coordinates for each diffraction disc. The alternating Si and SiGe layers provide distinct light and dark contrast, as shown in the virtual annular dark-field STEM image in Fig.~7. Strain maps $\varepsilon_{xx}$ and $\varepsilon_{yy}$ represent the out-of-plane and in-plane strain fields populated from two orthogonal reciprocal lattice vectors, $\langle 100 \rangle$
 and $\langle 110 \rangle$, respectively. 

The strain calculations are conducted by tracking the center coordinates of diffraction discs by our object detection model and relative displacements measured with respect to a reference region selected in the strain-free Si substrate. The oscillations in the colored map of strain field $\varepsilon_{xx}$  indicate positive (tensile) strain along the growth direction. The profile is plotted along the $\langle 100 \rangle$ direction, and the corresponding mean strain variation is approximately 0.8~\%, which is in good agreement with the expected maximum theoretical value in SiGe layers. 

The line profile analysis of $\varepsilon_{xx}$, shows uniform periodic strain modulations with relatively steep gradients at the boundaries between the Si and SiGe layers, accompanied by a top-flat maxima. This behavior is commensurate with the underlying Si/SiGe heterostructure and commonly observed in well-grown epitaxial materials. 

It is worth mentioning that the strain values reach down to minimum strain value of 0~\% across the alternating Si layers sandwiched between the SiGe layers, contributing to a well-defined experimental profile. The standard deviation representing the precision of the analysis is calculated using the Si substrate. The precision of 5x$10^{-4}$ is calculated across the strain-free Si substrate away from the thinnest part of the TEM sample. 
\begin{figure}[h]
	\centering
	\includegraphics[width=0.9\linewidth]{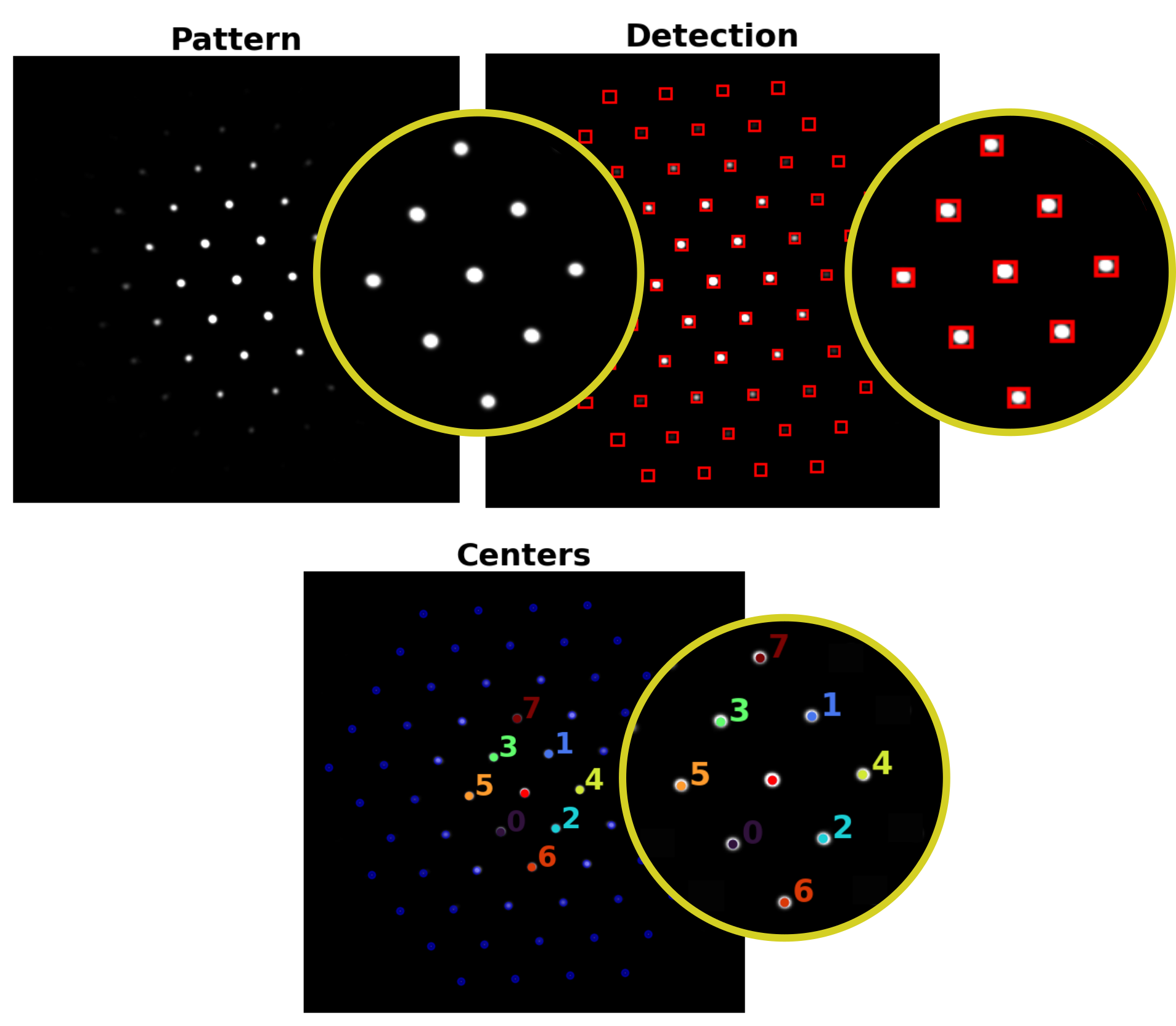}
	\caption{Top left to right: examples of electron diffraction patterns with Bragg discs outlined by bounding boxes as predicted by NDPD. Bottom: overlay showing the centers of the Bragg discs against the original diffraction pattern.}
	\label{fig:onecol_patterns}
\end{figure}

To obtain a parallel electron beam, and thereby small diffraction discs, in NBED STEM experiments, we tuned the microscope’s three condenser lens optics to achieve a convergence angle of 0.7~mrad \cite{yi2010flexible,zuo2004coherent}, resulting in a 2.6~nm probe size at full-width half maximum (FWHM). 

A careful analysis of the slope widths in the $\varepsilon_{xx}$ line profile reveals approximately 2.5~nm deviation from the theoretical low-amplitude strain modulation. This value agrees well with the expected spatial resolution at the interfaces, based on the experimental STEM probe size. 

The strain variations due to the bending of the thin TEM foil are also evident in the strain field maps. These variations contribute to the strain modulations on the order of 0.1-0.2~\% along the nominally strain-free $\langle 110 \rangle$ direction in the $\varepsilon_{yy}$ strain map.

Computational benchmarking was conducted on a desktop system equipped with an Intel i7 8-core CPU and a 12GB NVIDIA RTX 3060 GPU. To evaluate the end-to-end latency of the strain analysis, we utilized a 4D-STEM dataset acquired from an ROI scanned over a 97$\times$40~pixels field of view. A total of 3880 diffraction patterns, each with an image resolution of 512$\times$512~pixels, were processed asynchronously in 35~seconds, yielding an effective throughput of 110~fps. The model inference was carried out at an input resolution of 512$\times$512~pixels per pattern.
\begin{figure}[h]
	\centering
	\includegraphics[width=0.9\linewidth]{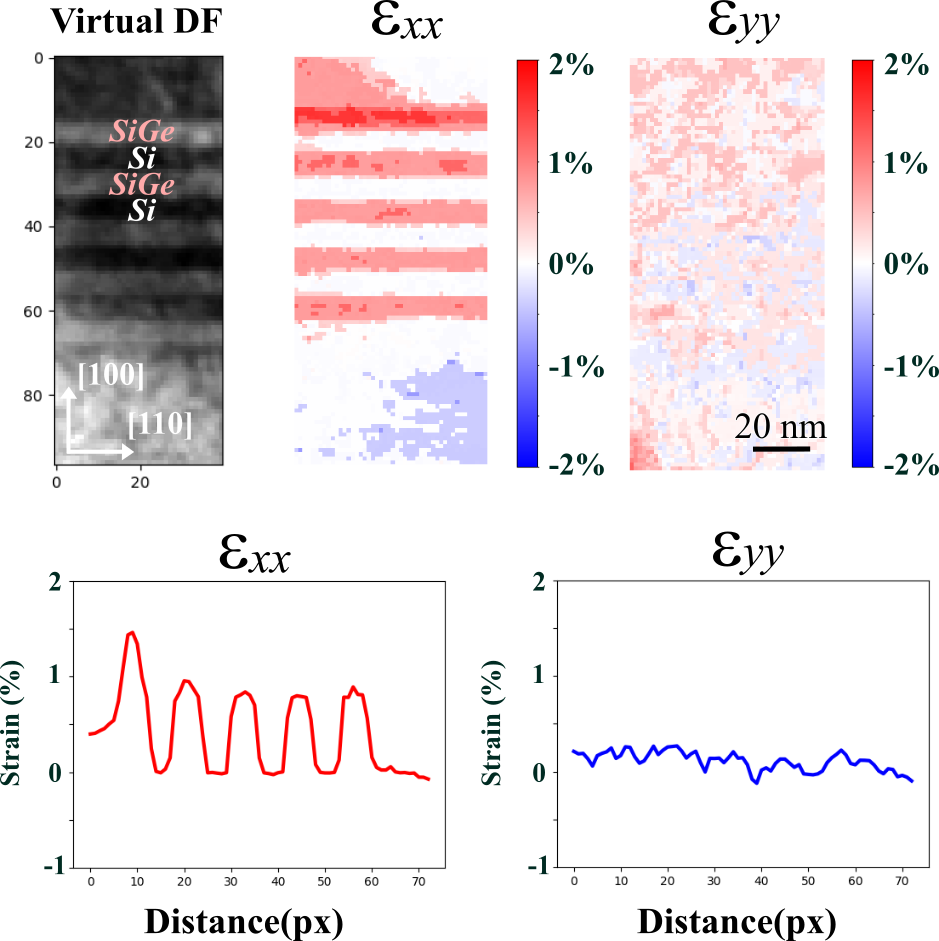}	
	\caption{Top left to right: the virtual DF image of Si/SiGe multilayers, the uniaxial strain map $\varepsilon_{xx}$ along the growth direction, and the orthogonal strain component $\varepsilon_{yy}$. Bottom left to right: the line profile across the thin-film growth $\langle 100 \rangle$ and strain-free $\langle 110 \rangle$ directions.}
	\label{fig:onecol_SiGe}
\end{figure}
\subsection{Phase and strain mapping of a multiphase metallic alloy}
The performance of our object detection-based peak localization workflow was evaluated on the task of disentangling complex interatomic displacement fields in a multiphase Ti-80Nb metallic alloy. These alloys frequently exhibit intricate phase distributions, often accompanied by characteristic displacement and strain fields from a multiphase microstructure. Conventional two-beam bright-field and dark-field imaging techniques are labor-intensive, dependent on user expertise, and frequently inadequate for resolving such complexity due to the nonlinear and overlapping contrast mechanism involved in phase formation.
 
Here, phase maps were generated by detecting and mapping interatomic lattice distances associated with each phase, analogous to composition-based XEDS phase mapping. Rather than relying on diffraction intensities, our approach reconstructs images based on the measured interatomic lattice distances. 

4D-STEM dataset for Ti-80Nb metallic alloy was acquired from a 128$\times$128~pixels ROI at a TEM foil thickness of 170 nm, with each electron diffraction pattern sampled at a resolution of 512$\times$512 pixels. A total of 16,384 diffraction patterns were processed using the asynchronous object detection workflow in 135~seconds, achieving an effective throughput of 121~fps with a batch size of 64 and an inference input resolution of 512x512 pixels per pattern. 
\begin{figure}[t]
	\centering
	\includegraphics[width=0.9\linewidth]{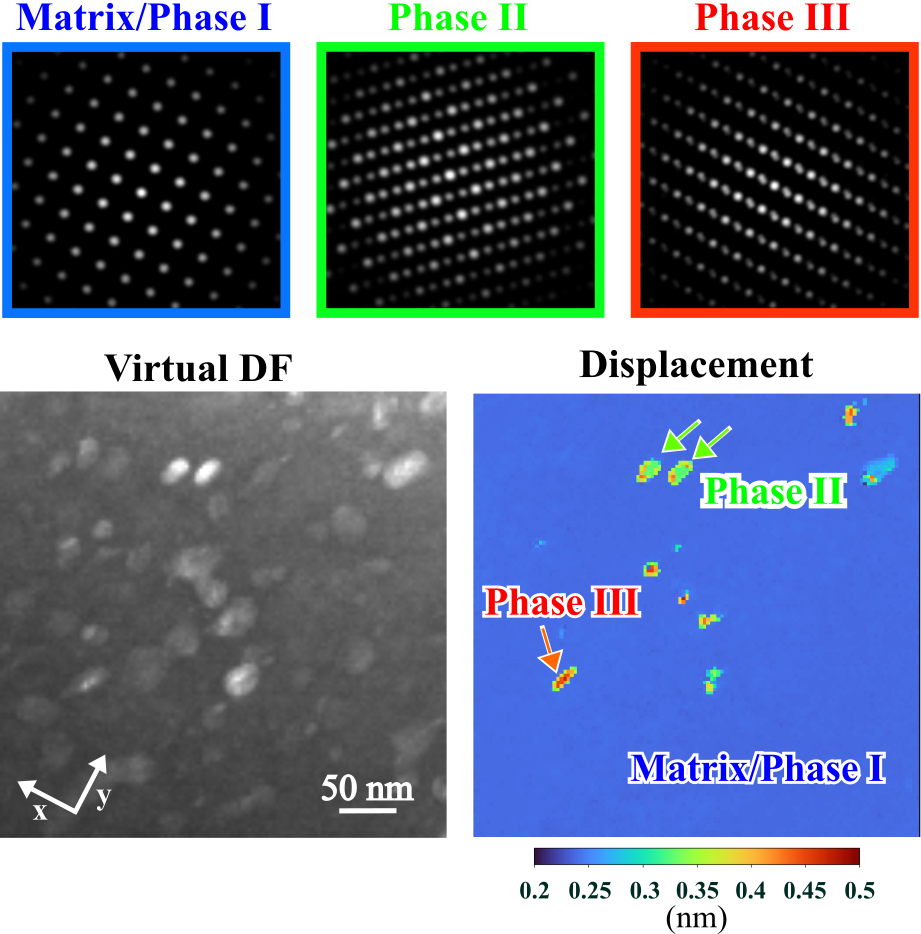}
	\caption{Top left to right: example electron diffraction patterns for matrix/coherent Phase I and incoherent Phases II and III overlapping with bcc matrix. bottom left to right: virtual DF image, and interatomic lattice displacement map.}
	\label{fig:onecol_TiNb}
\end{figure}

Fig.~8 presents example displacement field maps derived from two orthogonal $\langle 110 \rangle$ reciprocal lattice vectors, respectively. These maps reveal small, spherical bcc-precipitates (Phase I), approximately 50~nm in size, that are coherently embedded within the parent $\beta$-bcc matrix. In contrast, incoherent phases are identified as Phase II, corresponding to the ordered oxide $Ti_3O$, and Phase III, comprising  hcp-precipitates that overlap with the parent bcc-matrix, as evidenced by the electron diffraction patterns. 

The corresponding strain maps in Fig.~9 reveal a 1~\% tensile (positive) strain $\varepsilon_{xx}$, accompanied by a 0.5~\% compressive (negative) in $\varepsilon_{yy}$ and a 0.5~\% shear (negative) strain in $\varepsilon_{xy}$, all originating from the embedded bcc-precipitates. In addition to strain components, the rotation map indicates an approximate $1^\circ$ lattice rotation in the coherency strain field. These findings support the implementation of strain engineering strategies that can be leveraged to improve the mechanical performance of the alloy.
\begin{figure}[h]
	\centering
	\includegraphics[width=0.9\linewidth]{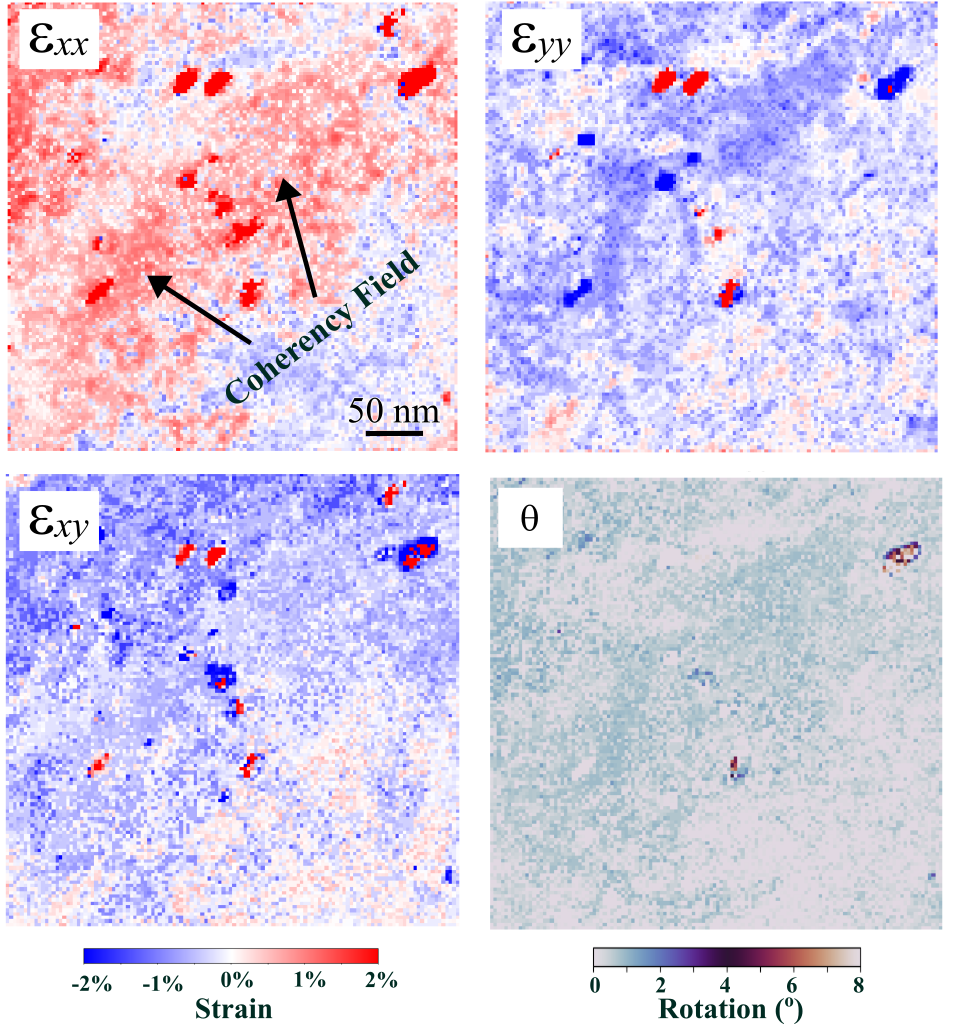}
	\caption{Top left to right: strain maps of $\varepsilon_{xx}$ and $\varepsilon_{yy}$.  Bottom left to right: strain map of $\varepsilon_{xy}$ and in-plane rotation map. }
	\label{fig:onecol_TiNbstrain}
\end{figure}

Currently, the model's performance is limited to the analysis of small Bragg disc patterns in NBED STEM datasets, which are consistent with those represented in the training dataset. To expand the applicability of our approach, we plan to examine the fidelity of object detection-based analysis of diffraction patterns in 4D-STEM datasets acquired under varying experimental conditions, including TEM foil thickness, Bragg disc size, and degree of disc overlap. This effort will require curating a more diverse and comprehensive dataset for model training.

\section{Conclusion}
In this work, we introduce a parallelized post-processing framework for large-scale 4D-STEM datasets, leveraging an efficient neural network-based object detection model. Rapid and accurate object detection presents a promising avenue for the analysis of electron diffraction patterns and the localization of interatomic lattice distances. Our approach enables sub-pixel precision in the detection of diffraction features, enabling high-accuracy crystallographic analysis. The ability to rapidly and reliably extract crystallographic information from 4D-STEM datasets acquired under varying experimental conditions accelerates materials discovery, from semiconductor development to alloy design.

\section{Data availability}
Python code and a demonstration video for the neural diffraction pattern detection workflow are available at \href{https://github.com/ArdaGen/4D-STEM-neural-diffraction-pattern-recognition-tempo4d/tree/main}
	{4D-STEM-neural-diffraction-pattern-recognition-tempo4d}.

\section{Acknowledgements}
The authors thank Andrew M. Thorn for acquiring the 4D-STEM datasets using the GATAN Stela hybrid-pixel electron detector and the STEMx system. Computational facilities (CNS-1725797) were provided by the Center for Scientific Computing (CSC), operated by the California NanoSystems Institute and the Materials Research Lab (MRSEC; NSF DMR 2308708) at UC Santa Barbara.

{
    \small
    \bibliographystyle{ieeenat_fullname}
    \bibliography{main}
}

\end{document}